\documentstyle[11pt,aaspp4]{article}

\def\about  {\hbox{$\sim$}}
\def\ga     {\mathrel{\hbox{\raise.3ex\hbox{$>$}\llap
                                {\lower.8ex\hbox{$\sim$}}}}}
\def\la     {\mathrel{\hbox{\raise.3ex\hbox{$<$}\llap
                                {\lower.8ex\hbox{$\sim$}}}}}
\def\x{\hbox{$\times$}}
\def\mic{\hbox{$\mu$m}}

\def\E#1{\hbox{$10^{#1}$}}

\def\kms{\hbox{km s$^{-1}$}}
\def\p{\phantom{0\,\,}}
\def\r{\phantom{0}}
\def\d{$^\circ$\,}
\def\HI  {\hbox{H\small I\ }}
\begin{document}
%
%
\rightline{\it Astrophysical Journal, in press}
\title{ 
        INFRARED SEARCH FOR YOUNG STARS \\ IN H{\large I} HIGH-VELOCITY CLOUDS}

\author{\large \v Zeljko Ivezi\'c}
\affil{Department of Astrophysical Sciences \\ Princeton University \\
Princeton, NJ 08544--1001 \\ E-mail: ivezic@astro.princeton.edu}

\author{\large  Dimitris M. Christodoulou}
\affil{Department of Physics and Astronomy \\ Louisiana State
University \\ Baton Rouge, LA 70803--4001 \\ E-mail:
dchristo@rouge.phys.lsu.edu}

%
%
\begin{abstract}

We have searched the IRAS Point Source Catalog and HIRES maps for
young stellar objects (YSOs) in the direction of five \HI
high-velocity clouds (HVCs). In agreement with optical searches in the
halo, no evidence was found for extensive star-forming activity inside
the high-latitude HVCs. Specifically, we have found no signs of star
formation or YSOs in the direction of the A IV cloud or in the
very-high-velocity clouds HVC~110-7-465 and HVC~114-10-440.  We have
identified only one young star in the direction of the M~I.1 cloud,
which shows almost perfect alignment with a knot of \HI emission.
Because of the small number of early-type stars observed in the halo,
the probability for such a positional coincidence is low; thus, this
young star appears to be physically associated with the M~I.1 cloud.
We have also identified a good YSO candidate in the \HI shell-like
structure observed in the core region of the low-latitude cloud
complex H (HVC~131+1-200).  This region could be a supernova remnant
with several other YSO candidates formed along the shock front
produced by the explosion.

In agreement with recent theoretical estimates, these results point to
a low but significant star-formation rate in intermediate and high
Galactic latitude HVCs. For M~I.1 in particular, we estimate that the
efficiency of the star-formation process is $M({\rm YSO})/M(\HI)\ga
10^{-4}-10^{-3}$ by mass.  Such efficiency is sufficient to account
for (a) the existence of the few young blue stars whose ages imply
that they were born in the Galactic halo, and (b) the nonprimordial
metallicities inferred for some HVCs if their metal content proves to
be low.

\end{abstract}


\newpage

%
\section                  {Introduction}
\label{intro}

Most Galactic neutral hydrogen (\HI) moves at velocities below about
100 \kms\ relative to the Local Standard of Rest (Heiles \& Habing
1974). Nevertheless, Muller, Oort, \& Raimond (1963) discovered \HI
clouds that move with velocities exceeding 100 \kms, too large to be
explained by differential Galactic rotation. Many of these
high-velocity clouds (HVCs) are located at intermediate and high
Galactic latitudes (Giovanelli, Verschuur, \& Cram 1973; Mathewson,
Schwarz, \& Murray 1977; Wakker 1991; Wakker \& van Woerden 1991; van
Woerden 1993) and do not appear to have any connection with the gas in
the Galactic disk. Their origin is still unclear, mainly because the
distances to the individual complexes are not known in most
cases. (See Wakker et al.  [1996] for a thorough review of HVC
formation scenarios following an attempt to constrain the distance of
two such clouds.)

Maps of the brightest HVC complexes have revealed the existence of
fine structure at 10 arcmin resolution, which was further resolved
into high-density cloud cores at 1 arcmin resolution (Wakker 1990;
Wakker \& Schwarz 1991, hereafter WS).  The \HI column densities in
these cores are estimated to be several times \E{20} cm$^{-2}$ and
their temperatures are generally between 30 and 300 K.  Such
conditions make the HVC cores possible sites of star formation.
Indeed, based on estimates of core collision timescales, core masses,
and Jeans masses, Dyson \& Hartquist (1983) predicted substantial
rates of star formation: up to 1000 early-type stars observable at any
time in about 10\% of all HVCs. In contrast, optical observations have
not detected any associated stars or star clusters. Unlike the dense
cores of molecular cloud complexes in the Galactic disk, the HVC cores
do not appear to be active sites of star formation.

Theoretical predictions for star formation rates in the HVCs have been
recently revised by Christodoulou, Tohline, \& Keenan (1997). Based on
the latest high-resolution radio observations, they estimate that HVCs
should not be sites of significant star formation. This result agrees
with the observations but poses yet another problem for the theory of
star formation: A number of young blue stars are observed at high
Galactic latitudes and at large distances from the Galactic disk
(Keenan et al. 1986, 1995; Conlon 1993; Little et al. 1995; Hambly et
al. 1996). One explanation is that these stars were formed in the disk
and were later ejected into the halo because of stellar dynamical
interactions in clusters, or because of supernovae explosions in
binaries (Leonard 1993 and references therein).  For about ten B-type
halo stars, however, the inferred ages are substantially shorter than
the travel times from the disk to their present locations, supporting
the alternative proposal that some stars do form in the Galactic halo.

This problem could be solved if there were star formation in HVCs even
at quite low rates.  Such rates would be in agreement with both
optical and \HI observations, and with the presence of just a few
unusual young stars at high Galactic latitudes.  In addition, low
rates would be in accord with the nonprimordial metallicities observed
in some HVCs (van Woerden 1993) and especially with metal abundances
whose origin cannot be attributed to alternative mechanisms (e.g.,
galactic fountains; Sembach 1995). These possibilities justify an
attempt to find observational evidence for star formation in HVCs.

Collapsing cloud cores and young stellar objects (YSOs) with gas
column densities of at least a few times \E{20} cm$^{-2}$ and standard
dust properties are detectable at infrared wavelengths by the IRAS
satellite (Boulanger, Baud, \& van Albada 1985). In fact, the spatial
resolutions of recent \HI observations (WS) and of IRAS maps are
comparable ($\sim 1$ arcmin), which greatly simplifies their
comparison. Wakker \& Boulanger (1986) searched for infrared emission
at 100 \mic\ in the direction of two HVCs, but this search produced a
negative result. They found no correlation between the \HI and 100
\mic\ emission, a result that they attributed to either very cold dust
($\la$ 20 K) or to a dust-to-gas ratio smaller than the standard
interstellar value. Wakker \& Boulanger did not consider IRAS maps at
other wavelengths, or information about YSO candidates available in
selected regions from the IRAS Point Source Catalog (PSC). In this
work we have extended the search to five HVCs (A~IV, M~I.1, complex~H,
HVC~110-7-465, and HVC~114-10-440) recently observed in \HI by WS,
utilizing both the PSC and HIRES maps (HIgh RESolution IRAS images) at
all four IRAS wavelengths (12, 25, 60, and 100 \mic). Our searching
procedure and the results for each HVC are described in \S~2. In \S~3
we summarize our findings.

\section{     The Search for YSOs in the IRAS Database    }
\label{sources}

\subsection     {The Search in the IRAS PSC      }

WS observed the \HI emission from several HVCs with 1 arcmin spatial
resolution and 1 \kms\ velocity resolution. We have used their maps to
positionally constrain our search for YSOs. The covered areas are the
smallest rectangles which fully include the regions of observed \HI
emission.  Their typical sizes are about $1^\circ$.  The first step
was to find all IRAS point sources in the corresponding directions;
this search resulted in 21 sources. For each source, we can
distinguish whether it is a YSO or a late-type star by using the
infrared colors based on their four IRAS fluxes: Late-type stars are
much brighter at the two shorter wavelengths than at the two longer
wavelengths, while the opposite is true for young stars, star-forming
regions, and galaxies (Ivezi\' c \& Elitzur 1996). However, this
criterion can be safely applied only if the source has good-quality
flux measurements (quality 3; see IRAS Explanatory Supplement). In
addition to excluding all bona fide late-type stars, we also
eliminated all sources that do not show positional coincidence with
the structures seen in \HI emission. The remaining 8 YSO candidates
are listed in Table 1. For these sources, we have obtained HIRES
images at all four IRAS wavelengths. The location of one additional IR
source (not listed in Table 1 since it is not in the PSC) in the M~I.1
cloud and the locations of three sources in complex H are also marked
in Fig.~1 that depicts the corresponding \HI emission maps
kindly supplied by B. Wakker. These sources are the best candidates
for embedded YSOs in the corresponding clouds.  We proceed to discuss
the results for each cloud separately.

\begin{table}[t]
\begin{center}
\begin{minipage}{7in}
\caption{}
\begin{tabular}{ccccccccl}
\multicolumn{9}{c}{\sc IRAS PSC YSO Candidates Toward \HI HVCs} \\
\hline\hline
     HVC    & IRAS Source&       RA       &      Dec      & F$_{12}$    & F$_{25}$    & F$_{60}$    & F$_{100}$   & YSO? \\
     (1)    &     (2)    &       (3)      &      (4)      & (5)    & (6)    & (7)    & (8)    & (9)  \\
\hline\hline
M~I.1       & 11235+4300 & 11$^h$ 23$^m$ 31$^s$&+43\d \r0$^\prime$\p0$^{\prime\prime}$&0.27 (3)&0.25 (1)&0.53 (1)&1.18 (1)& No \\
A~IV        & 09059+6227 &\r9$^h$\p5$^m$ 57$^s$&+62\d  27$^\prime$\p4$^{\prime\prime}$&0.25 (1)&0.25 (1)&1.71 (3)&4.99 (3)&  No \\
H           & 01572+6248 &\r1$^h$ 57$^m$ 14$^s$&+62\d  48$^\prime$\p1$^{\prime\prime}$&0.25 (1)&0.25 (1)&0.64 (1)&5.55 (1)&  Yes \\
H           & 01588+6219 &\r1$^h$ 58$^m$ 50$^s$&+62\d  19$^\prime$ 30$^{\prime\prime}$&0.26 (3)&0.25 (1)&0.57 (1)&24.1 (1)&  Yes \\
H           & 02020+6247 &\r2$^h$\p2$^m$\p1$^s$&+62\d  47$^\prime$ 40$^{\prime\prime}$&0.37 (1)&0.25 (1)&1.66 (3)&5.84 (3)&  Yes \\
H           & 02021+6228 &\r2$^h$\p2$^m$ 11$^s$&+62\d  28$^\prime$ 12$^{\prime\prime}$&0.38 (3)&0.25 (1)&0.40 (1)&5.11 (1)&  Yes \\
H           & 02022+6236 &\r2$^h$\p2$^m$ 16$^s$&+62\d  36$^\prime$ 10$^{\prime\prime}$&0.69 (3)&0.26 (1)&0.45 (1)&6.33 (1)&  Yes \\
114-10-440  & 23486+5103 & 23$^h$ 48$^m$ 38$^s$&+51\d \r3$^\prime$ 45$^{\prime\prime}$&0.41 (3)&0.28 (1)&0.40 (1)&1.45 (1)& No \\
\hline\hline
\multicolumn{9}{l}{(1) HVC name.}\\
\multicolumn{9}{l}{(2) IRAS source name.}\\
\multicolumn{9}{l}{(3,4) Right Ascension and Declination (epoch 1950).} \\
\multicolumn{9}{l}{(5-8) Fluxes in Jy at the four IRAS wavelengths;  flux qualities
are listed in parentheses} \\
\multicolumn{9}{l}{\p \p \p (3: very good, 2: good, 1: upper limit).} \\
\multicolumn{9}{l}{(9) Our best determination of the YSO nature for each source, based on analysis of HIRES images.}\\

\end{tabular}
\end{minipage}
\end{center}
\end{table}

\subsection               {Cloud M I.1}

Cloud M I.1, one of the two brightest regions in complex M, contains
five embedded \HI cores (Fig. 1c in WS).  An upper limit to the
distance of the neighboring M~II-M~III region, $d\approx 5$ kpc, has
been determined from absorption measurements in the direction of a
background star (Danly, Albert, \& Kuntz 1993; Keenan et al. 1995).
We found one point source, IRAS 11235+4300, in the direction of M~I.1.
The four HIRES maps, centered on this source, are shown in
Fig.~2. IRAS 11235+4300 (HD~99327) is barely visible in the
12 and 25 \mic\ maps and nonexistent at 60 and 100 \mic, indicating
that it is a late-type star, an identification also supported by its
spectral type K0.  The large fluxes at 60 and 100
\mic\ listed for this source in Table 1 are probably caused by
imperfections in the correction procedure for the background cirrus
emission (Ivezi\'c \& Elitzur 1995).

On the other hand, there is in Fig.~2 a sign of point-like
very cold emission indicative of a YSO at RA=11$^h$ 23$^m 51\pm2^s$
and Dec=43$^\circ 21.3\pm0.5'$ corresponding to the position just
below the upper edge, slightly shifted to the left from the image
center. This emission appears as a faint source at 60 \mic\ (just
above the noise level) and as a quite bright knot at 100
\mic. Interestingly, the knot coincides precisely with a peak of \HI
emission (Fig.~1a). QSO~1123+434 (RA=11$^h$ 23$^m 49.41^s$,
Dec=43$^\circ 26' 7.4''$) is in the vicinity of the knot as well, but
it does not coincide with the region of the infrared emission.

Assuming a cloud with constant temperature $T$ and standard dust
optical properties, the temperature of the emitting dust can be
estimated from the ratio F$_{100}/{\rm F}_{60}$ of the 100 to 60 \mic\
fluxes that is obtained from the corresponding images. For the bright
knot at 100 \mic, we find that F$_{100}/{\rm F}_{60}\approx 2.5$ and
$T=96/\ln{(10 {\rm F}_{100}/{\rm F}_{60})} \approx 30$ K.  For a
standard dust-to-gas ratio and measured total \HI column density of
${\cal N}$ = 4.36\x\E{20} cm$^{-2}$ (core \#4 in Table 2 of WS; see
also their Table 1), the intensity of the resolved emission at 100
\mic\ would be $I_{100}=4\x 10^{-17} ({\cal N}/{\rm cm}^{-2})
\exp{(-144/T)} \approx 140$ MJy sr$^{-1}$. Since the measured peak
intensity is about 1 MJy sr$^{-1}$, and the size of the IRAS point
spread function (PSF) at 100 \mic\ is \about 1 arcmin, the angular
size of the emitting region is estimated to be about 5 arcsec.
Assuming an upper limit of 5 kpc for the cloud distance, this angular
size corresponds to an upper limit of \E{17} cm, comparable to
expected collapsing-core sizes. Thus, it is conceivable that the
source at RA=11$^h$ 23$^m 51\pm2^s$ and Dec=43$^\circ 21.3\pm0.5'$ is
a YSO physically associated with cloud core \#4 in HVC M~I.1.

\subsection                {Cloud A IV}

Cloud A IV is the brightest region in complex A (Wakker \& van Woerden
1991) and shows filamentary structures in \HI (Fig. 1a in WS).  We
found one point source in this region, IRAS 09059+6227, that can be
identified in all four panels shown in Fig.~3. Its brightness
increases rapidly with wavelength indicating that the source is not a
late-type star. Cross-referencing of various catalogs showed that this
source positionally coincides with the galaxy UGC~4803 (RA=9$^h$ 5$^m
56.9^s$, Dec=62$^\circ 26' 58''$), also observed by Davis \& Seaquist
(1983). It seems that IRAS 09059+6227 is a galaxy projected on the sky
near the edge of A~IV.

\subsection                 {Complex H}

Unlike complexes A and M, the extended complex H (diameter 20$^\circ$)
is centered near the Galactic plane. The large-scale velocity field
suggests the presence of an expanding shell with velocity $v=70$ \kms
~(Fig. 1i in Wakker \& van Woerden 1991). The origin of this cloud is
uncertain. WS discuss the possibility of a dense high-velocity cloud
impacting the disk from the side, creating a shock front.
Alternatively, the expanding shell geometry could be indicative of a
supernova remnant.

Table 1 lists five sources that could be associated with the core of
complex H (HVC~131+1-200), although only three of them (IRAS
01572+6248, 01588+6219, and 02021+6228; Fig.~1b) show good
positional coincidence with the \HI shell structure observed in the
central region of the complex (see also Fig.~1e in WS).  IRAS
01572+6248 is located at the center of the 60 and 100 \mic\ panels
shown in Fig.~4.  Part of the shell outline can be seen in
the lower half of the same panels, and its position and size roughly
correspond to the inner shell observed in \HI emission. Note that
there is an offset between the center positions: from IRAS maps we get
RA=1$^h$ 58$^m$ and Dec=62$^\circ 31'$, while the center position from
the \HI maps is RA=1$^h$ 58$^m$ 30$^s$ and Dec=62$^\circ 34'$. The
difference is about 3 arcmin or 3 FWHM of the PSF.

The close correspondence between the observed \HI shell structure and
the infrared maps makes the association of IRAS 01572+6248 with
complex H very likely.  Emission from this source increases toward
long wavelengths indicating a YSO. Several other blobs can be seen
along the shell's edge in the 100 \mic\ map and their emission also
increases with wavelength.  Unfortunately, further analysis that could
provide more clues about their nature is hampered by their weak
emission which is not much stronger than the noise level. Thus, the
detection of these sources remains tentative.

We have repeated the analysis of \S~2.2 for the three sources whose
positions coincide with the \HI shell in the core of complex H
(Fig.~1b). In all cases, we find from the images that F$_{100}/{\rm
F}_{60}\approx 2.0$ implying a temperature for the emitting dust of
about 32 K. For a standard dust-to-gas ratio and for total \HI column
densities of ${\cal N}$ = (1.3-2.6)\x\E{20} cm$^{-2}$, the intensities
of the resolved emission at 100 \mic\ would range between 60 and 120
MJy sr$^{-1}$. Since the measured peak intensities are between 3 and 6
MJy sr$^{-1}$, respectively, the angular sizes of the emitting regions
are all estimated to be about 0.22 arcmin. With a lower limit on the
cloud distance of about 5 kpc (Wakker \& van Woerden 1997), we derive
a lower limit of 3\x\E{17} cm for the sizes of these regions.  This
value is at least three times larger than the size of the YSO found in
cloud M I.1. A more precise comparison cannot be made because of the
uncertainty in the determinations of the distances of the two HVCs.

\subsection      {HVC 110-7-465 and HVC 114-10-440}

These two very-high-velocity clouds are only 5$^\circ$ apart in the
sky and have similar structures (Figs. 1f and 1h in WS). Well-defined
gradients in the velocity fields indicate that both clouds are
rotating.  In the region outlined by \HI emission from these clouds,
we found only one source, IRAS 23486+5103. It has a very good
positional coincidence with the peak of the \HI emission in HVC
114-10-440, and at first it appeared to be the best candidate for a
YSO associated with an HVC. Unfortunately, inspection of HIRES maps
shows that this source is a late-type star (HD~223626) because the
intensity of its emission decreases with wavelength, and its spectral
type is listed as K2.  An increase in fluxes listed in Table 1 from 
60 to 100 \mic\ is probably due to imperfections in the correction 
procedure for the background cirrus emission, since it is not corroborated 
by the detailed images.

HIRES maps of the region toward these two clouds show hints of two
more YSOs based on their stronger emission at 60 and 100 \mic\ relative
to the shorter wavelengths. These sources are located quite far from
the regions of elevated \HI column density and their apparent association
with these HVCs is probably just a projection effect. Thus, our search for
embedded YSOs has produced negative results in the cases of HVC 110-7-465
and HVC 114-10-440.

\section            {Summary and Conclusions}
\label{summary}

We have examined the infrared emission in the direction of five HVCs,
searching the IRAS Point Source Catalog and the corresponding
high-resolution images at all four IRAS wavelengths. We have found no
evidence for extensive star-formation activity in any of the surveyed
regions that show \HI emission (WS). Yet, a low rate of star formation
inside HVCs remains a possibility in light of the discovery of a few
individual early-type stars.  Specifically, one such star was found in
the direction of the M~I.1 cloud and it is quite possible that this
star is embedded in the cloud (\S~2.2). In contrast, no early-type
stars were found toward the A~IV cloud (\S~2.3). The remaining five
sources (one likely detection and four tentative YSO candidates) were
found toward the inner shell structure in the low-latitude complex H
(HVC~131+1-200) and they could have formed in the shock front created
by a supernova explosion (\S~2.4). Finally, no YSOs were found toward
the very-high-velocity clouds HVC 110-7-465 and HVC 114-10-440
(\S~2.5).

A low star-formation rate at intermediate and high Galactic latitudes,
consistent with the single YSO found in M~I.1, is in agreement with
recent theoretical estimates (Christodoulou et al. 1997), showing that
it is very difficult for a collisional mode of star formation to be
excited in halo HVCs. The star-formation efficiency in M~I.1 can be
estimated as the fraction of the cloud's mass that is converted into a
star: Using $d\leq 5$ kpc for the distance and ${\cal N}\approx$
3\x\E{20} cm$^{-2}$ for the \HI column density of the densest cloud
core with angular diameter $1^\circ$, we find from the equations of
Christodoulou et al. (1997) a radius of $R\la 44$ pc and a mass of
$M(\HI)\la 10^4$ M$_\odot$.  Assuming then a mass of $M({\rm YSO})\sim 1-10$
M$_\odot$ for the detected YSO, we find that $M({\rm YSO})/M(\HI)\ga
10^{-4}-10^{-3}$.  Such efficiency can account for the observed
nonprimordial metallicities (e.g., van Woerden 1993) only if they
prove to be low and for the existence of $\sim 10$ young blue stars
whose ages imply that they were born in the Galactic halo (e.g.,
Conlon 1993). These halo stars are not clearly projected toward HVC
cores or regions of pronounced \HI emission, and to further support
the latter hypothesis it is important to examine whether they have had
enough time to abandon their HVC birthplaces.

\acknowledgments
We are grateful to C. Brogan for her expert help with the astronomical
image viewing program Aipsview developed at NCSA and to J. Knapp for a
careful reading of the manuscript and for her suggestions. We also
thank B. Wakker for sending us his \HI emission maps and an anonymous
referee for critical comments that helped us improve the paper. This
work has made use of the SIMBAD database and was supported in part by
NASA grant NAG5--2777, NSF grant AST 95--28424, and by the Center for
Computational Sciences of the University of Kentucky.
\clearpage

%


\clearpage

\section*{FIGURE CAPTIONS}

\figcaption{The locations of four YSO candidates probably embedded
in the corresponding HVCs. The gray scale images are \HI maps obtained
by WS. A bar denotes an angular scale of 10 arcmin for both maps and
the positions of the sources are indicated by open circles.  (a)~The
YSO in M~I.1 at RA=11$^h$ 23$^m 51^s$ and Dec=43$^\circ 21.3'$ (see
\S~2.2).  (b)~Three IRAS PSC sources in complex H (see \S~2.4).
\label{fig0}
}

\figcaption{HIRES images at four IRAS wavelengths (12, 25, 60, and 100 \mic)
in the direction of cloud M~I.1. The $1^\circ \times 1^\circ$ images
are centered on the source IRAS 11235+4300 at the coordinates listed
in Table 1.  The vertical bar next to the right edge of each panel
shows the intensity scale in MJy sr$^{-1}$. The upper cut-off in each
panel was set to 2 MJy sr$^{-1}$ in order to emphasize the regions
with low-level intensities.
\label{fig1}
}

\figcaption{As in Fig. 1, but in the direction of cloud A~IV.
The $1^\circ \times 1^\circ$ images are centered on the source IRAS
09059+6227 at the coordinates listed in Table 1.
\label{fig2}
}

\figcaption{As in Fig. 1, but in the direction of complex H.
The $1^\circ \times 1^\circ$ images are centered on the source IRAS
01572+6248 at the coordinates listed in Table 1.
\label{fig3}
}

\end{document}